\input harvmac
\input epsf
\noblackbox
\def\npb#1#2#3{{\it Nucl.\ Phys.} {\bf B#1} (#2) #3}
\def\plb#1#2#3{{\it Phys.\ Lett.} {\bf B#1} (#2) #3}
\def\prl#1#2#3{{\it Phys.\ Rev.\ Lett.} {\bf #1} (#2) #3}
\def\prb#1#2#3{{\it Phys.\ Rev.} {\bf B#1} (#2) #3}
\def\prd#1#2#3{{\it Phys.\ Rev.} {\bf D#1} (#2) #3}

\def\cmp#1#2#3{{\it Commun.\ Math.\ Phys.} {\bf #1} (#2) #3}
\def\jgp#1#2#3{{\it J. Geom.\ Phys.} {\bf #1} (#2) #3}

\def\jhep#1#2#3{{\it JHEP\/} {\bf #1} (#2) #3}


\def\bberkeley{\centerline{\it Center for Theoretical Physics and 
Department of Physics, University of California}
\centerline{\it Berkeley, CA 94720-7300, USA}
\centerline{\it and}
\centerline{\it Theoretical Physics Group, Lawrence Berkeley National 
Laboratory}
\centerline{\it Berkeley, CA 94720-8162, USA}}

\font\authfont=cmr12

\def\frac#1#2{{#1 \over #2}}
\def\bra#1{{\langle #1 |}}
\def\ket#1{{| #1 \rangle}}

\def\semi{\subset\kern-1em\times\;}

                   \def\CG{{\cal G}}
\def\CH{{\cal H}}

\def\R{{\bf R}}

\font\mathbb msbm7 at 10pt

\def\ZZ{\hbox{\mathbb Z}}

\def\SU2#1{{\widehat{SU(2)}_{#1}}}

\def\insfig#1#2#3#4{\midinsert\vbox to #4truein{\vfil\centerline{
\epsfysize=#4truein\epsfbox{#3}}}
\narrower\narrower\noindent{\bf #1.} #2\endinsert}

\Title{\vbox{\baselineskip12pt
\hbox{hep-th/0501233}
\hbox{\ }}}
{\vbox{\centerline{Boundary Scattering in 1+1 Dimensions}
\bigskip
\centerline{as an Aharonov-Bohm Effect}}}

\nref\cardy{J. Cardy, 
``Conformal Invariance and Surface Critical Behavior,'' \npb{240}{1984}{514}.}

\nref\affleckA{I. Affleck, 
``Conformal Field Theory Approach to Quantum Impurity Problems,'' cond-mat/9311054.}

\nref\affleckB{I. Affleck and A.W.W. Ludwig, 
``Exact Conformal Field Theory Results on the Multichannel Kondo Effect: Single-fermion Green's Function, Self-energy, and resistivity,'' 
\prb{48}{1993}{7297}.}

\nref\callanmon{C. G. Callan and S. R. Das,
``Boundary Conditions on the Monopole-Dirac Equation,''
\prl{51}{1983}{1155}.}

\nref\kane{C. L. Kane and M. P. A. Fisher, 
``Transport in a One-channel Luttinger Liquid,''
\prl{68}{1992}{1220}.}

\nref\horava{P. Horava, ``Strings On World-Sheet Orbifolds,''
Nucl. Phys. {\bf B327} (1989) 461, ``Background Duality Of Open
String Models,'' Phys. Lett. {\bf B321} (1989) 251.}

\nref\oldpolch{J. Dai, R. G. Leigh, and J. Polchinski,
``New Connections Between String Theories,'' Mod. Phys. Lett. {\bf A4}
(1989) 2073.}
\nref\otherpolch{J. Polchinski, ``Combinatorics of Boundaries In
String Theory,'' Phys. Rev. {\bf D50} (1994) 6041, hep-th/9407031.}

\nref\sen{A. Sen,
``Rolling Tachyon,'' \jhep{0207}{2002}{65}, hep-th/0203211.}

\nref\petrorb{P. Ho\v rava, ``Chern-Simons Gauge Theory on Orbifolds: Open 
Strings from Three Dimensions,'' \jgp{21}{1996}{1}, hep-th/9404101.}

\nref\witten{E. Witten, 
``Quantum Field Theory and the Jones Polynomial,'' \cmp{121}{1989}{351}.}

\nref\elitzur{S. Elitzur, G. Moore, A. Schwimmer, and N. Seiberg,  
``Remarks on the Canonical Quantization of the Chern-Simons-Witten Theory,'' 
\npb{326}{1989}{108}.}

\nref\callan{C.G. Callan, I.R. Klebanov, A.W.W. Ludwig and J.M. Maldacena, 
``Exact Solution of a Boundary Conformal Field Theory,''\npb{422}{1994}{417}, 
hep-th/9402113;\ see also:\  
J. Polchinski and L. Thorlacius, 
``Free Fermion Representation of a Boundary Conformal Field Theory,'' 
\prd{50}{1994}{622}, hep-th/9404008.}

\nref\MSZoo{G. Moore and N. Seiberg, ``Taming the Conformal Zoo,''
\plb{220}{89}{422}}

\nref\Kogan{Ya.I. Kogan, ``The Off-Shell Closed Strings as Topological Open
Membranes.  Dynamical Transmutation of World-Sheet Dimension,''
\plb{231}{89}{377}.}

\nref\CarlipKogan{S. Carlip and Ya.\ Kogan, ``Quantum Geometrodynamics of the
Open Topological Membrane and String Moduli Space,'' \prl{64}{90}{1487}.}

\nref\DijkW{R. Dijkgraaf and E. Witten, ``Topological Gauge Theories and Group
Cohomology,'' \cmp{129}{90}{393}.}

\nref\FFFSA{G. Felder, J. Fr\"ohlich, J. Fuchs and C. Schweigert,
``Conformal Boundary Conditions and Three-dimensional Topological Field Theory,''
\prl{84}{2000}{1659}, hep-th/9909140.}

\nref\FFFSB{G. Felder, J. Fr\"ohlich, J. Fuchs and C. Schweigert,
``Correlation Functions and Boundary Conditions in RCFT and Three-dimensional Topology,''
Compos.Math. {\bf 131} (2002) 189, hep-th/9912239.}

\nref\Cardy{J.L. Cardy, ``Boundary Conditions, Fusion Rules, and the Verlinde
Formula,'' \npb{324}{89}{581}.}

\nref\Gab{M.R. Gaberdiel, A. Recknagel and G. M. T. Watts,
``The Conformal Boundary States for SU(2) at Level 1,'' \npb{626}{2002}{344},
  hep-th/0108102.}

\nref\Ferreira{P. C. Ferreira, I. I. Kogan and B. Tekin,
``Toroidial Compactification in String Theory from Chern-Simons Theory,''
\npb{589}{2000}{167}, hep-th/0004078.}


\bigskip
\centerline{\authfont Surya Ganguli, Petr Ho\v rava and Anthony Ndirango}
\medskip\bigskip\medskip
\baselineskip14pt
\bberkeley
\centerline{\tt sganguli, horava, andirang@socrates.berkeley.edu}
\medskip\bigskip\medskip
\centerline{\bf Abstract}
\bigskip
\noindent

The boundary scattering problem in $1+1$ dimensional CFT is relevant
to a multitude of areas of physics, ranging from the Kondo effect in
condensed matter theory to tachyon condensation in string theory.
Invoking a correspondence between CFT on $1+1$ dimensional manifolds
with boundaries and Chern-Simons gauge theory on $2+1$ dimensional
$\ZZ_2$ orbifolds, we show that the $1+1$ dimensional conformal
boundary scattering problem can be reformulated as an Aharonov-Bohm
effect experienced by chiral edge states moving on a $1+1$ dimensional
boundary of the corresponding $2+1$ dimensional Chern-Simons system.
The secretly topological origin of this physics leads to a new and
simple derivation of the scattering of a massless scalar field on the
line interacting with a sinusoidal boundary potential.  
\Date{January 2005}
\newsec{Introduction}

\par Conformal field theory (CFT) in $1+1$ dimensions on a manifold with
boundary \refs{\cardy} serves as a unifying framework for many
important problems in physics.  Examples in both condensed matter and
string theory include quantum impurities in metals (the Kondo effect
\refs{\affleckA,\affleckB}), proton-monopole scattering
\refs{\callanmon}, quantum wire junctions \refs{\kane}, D-branes
\refs{\horava-\otherpolch}, and tachyon condensation \refs{\sen}.  In
all of these problems the scattering amplitude of a bulk excitation in
the presence of various boundary conditions is of physical interest.
In this letter we prove a conjecture first stated in \refs{\petrorb}
and show that this amplitude is essentially an Aharanov-Bohm phase by
exploiting a connection to Chern-Simons (CS) theory.  Thus all of the
physical phenomena above are secretly topological in origin.

\par
In section 2 we briefly review the connection between $2+1$
dimensional CS theory and CFT in $1+1$ dimensions \refs{\witten, \elitzur}, and an
extension of this relation connecting boundary CFT to CS theory
on a $\ZZ_2$ orbifold \refs{\petrorb}.  In particular we explain the $2+1$ dimensional
origin of the bulk one point function in the presence of a boundary.  In
section 3 we state the formulation of Kondo scattering in terms of
boundary CFT, and show that the scattering amplitude is precisely an
Aharanov-Bohm phase when interpreted using the results of section 2.
In section 4 we use similar techniques applied to the boundary CFT of
a massless scalar field interacting with a sinusoidal boundary
potential, which enables us to give a new and simpler topological derivation of
some results in \refs{\callan}.  We conclude in section 5.

\newsec{Boundary CFT and CS Theory on Orbifolds}

\subsec{The Chiral Picture}
\par 
CS theory with a compact, simple gauge group $\cal G$ on a three manifold $M$ has
the action 
\eqn\eeaction{
S[A]=\frac{k}{4\pi}\int_{M}\Tr{(A \wedge dA + \frac{2}{3}A\wedge A \wedge A)}. 
}
The observables of the theory are products of Wilson lines
\eqn\eeobs{W_R(C)={\rm Tr}_R\,P\exp \int _CA,}
where $R$ belongs to the finite set of integrable representations of the
Kac-Moody algebra $\widehat\CG$ at level $k$, and $C$ is a closed line in
$M$.  In the case where $M = \Sigma  \times \R$ with Wilson lines 
in representations $R_i$ piercing $\Sigma$ at points $z_i$, one can 
canonically quantize the theory \refs{\witten,\elitzur} to obtain
a finite-dimensional quantum Hilbert space $\CH_{\Sigma,R_i}$.  The connection
to CFT lies precisely in the fact that $\CH_{\Sigma,R_i}$ is also the
space of conformal blocks of a $2D$ rational CFT, namely a level $k$ WZW model
associated with group $\cal G$ \refs{\witten,\elitzur}.  Thus the chiral
correlators of primary fields in the CFT can be thought of as elements of
the CS quantum Hilbert space.

\par 
Furthermore, one can use the path integral formulation of CS theory to
pick out specific elements of $\CH_{\Sigma,R_i}$.  Suppose $M$ is now
a three manifold with boundary $\Sigma$ and appropriate Wilson line
insertions ending on the boundary.  To define the path integral on
$M$, one must fix the gauge field $A$ over the boundary $\Sigma$.  The
path integral over $A$ in the bulk is then a function (or more
precisely a section of a line bundle) over the space of gauge
equivalence classes of $A$ on the boundary $\Sigma$.  However
$\CH_{\Sigma,R_i}$ itself can be thought of as the space of such
sections, and so the path integral on $M$ picks out a distinguished
element of this space.

\par
In the important case where $M$ is a solid torus with boundary
$\Sigma = T^2$, $\CH_{\Sigma}$ is spanned by the basis
$\ket{\Psi_{T^2}^{R_i}}$ where again $R_i$ is an integrable
representation of $\widehat\CG$ at level $k$.  In particular,
$\ket{\Psi_{T^2}^{R_i}}$ can be obtained by evaluating the path integral on
$M$ with a single Wilson loop in the representation $R_i$ inserted along
the unique noncontractible cycle of $M$.  Since $SL(2;\ZZ)$ acts as
the group of homotopy classes of diffeomorphisms of $T^2$, it induces
an action on $\CH_{T^2}$.  In the following, we will need the matrix
elements of the generator \eqn\eeS{S = \left(\matrix{0 & -1\cr 1 &
0\cr }\right) \in SL(2;\ZZ)} in the basis $\ket{\Psi_{T^2}^{R_i}}$.
These elements can be read off from the transformation properties of
the characters of $\widehat\CG$ at level $k$ under the modular
transformation $\tau \rightarrow -1/\tau$ where $\tau$ is the complex
structure modulus of $T^2$.  When $\widehat\CG = \widehat{SU(2)}$,
there are $k+1$ integrable representations with highest weight
$\lambda$ where $0 \leq \lambda \leq k$, and in this basis, the matrix
elements of the corresponding modular S-matrix are given by,
\eqn\eemodS{S_{\lambda\mu} \equiv \bra{\Psi_{T^2}^{\lambda}} S
\ket{\Psi_{T^2}^{\mu}} =
\sqrt{\frac{2}{k+2}}\sin{\frac{\pi(\lambda+1)(\mu+1)}{k+2}}.  }

\subsec{Left Movers and Right Movers}
\par 
So far we have reviewed the three dimensional origin of chiral
correlators in the CFT.  Now, to explain the origin of boundary CFT, we
must first see how both chiral halves consisting of left-movers and
right-movers come from three dimensions in the case without boundary.
This issue was addressed in \refs{\elitzur,\MSZoo-\CarlipKogan}.  The
key idea is to consider a thickening $M = \Sigma \times [0,1]$ of the
closed Riemann surface $\Sigma$.  One can then show \refs{\elitzur}
that the gauge invariant degrees of freedom of CS theory live on the
two disjoint boundary components of $M$, and correspond to the left-movers 
and right-movers of the full CFT.  The insertion of a
left-moving primary field in representation $R_i$ at point $z_i$ in
the CFT corresponds, in CS theory, to the insertion of an oriented
Wilson line in the representation $R_i$ travelling from the left to right
boundary components, along $z_i \times I$.  For the insertion of
a right-moving primary, the orientation of the Wilson line is
reversed, as in figure~(1.a).
\insfig{1}{Elements of the space of two point conformal blocks
on $\Sigma$.  (a) $M = \Sigma \times I$ with Wilson line insertions.  The
shaded boundary represents two disjoint copies of $\Sigma$. (b) A standard
basis of the space of conformal blocks.}{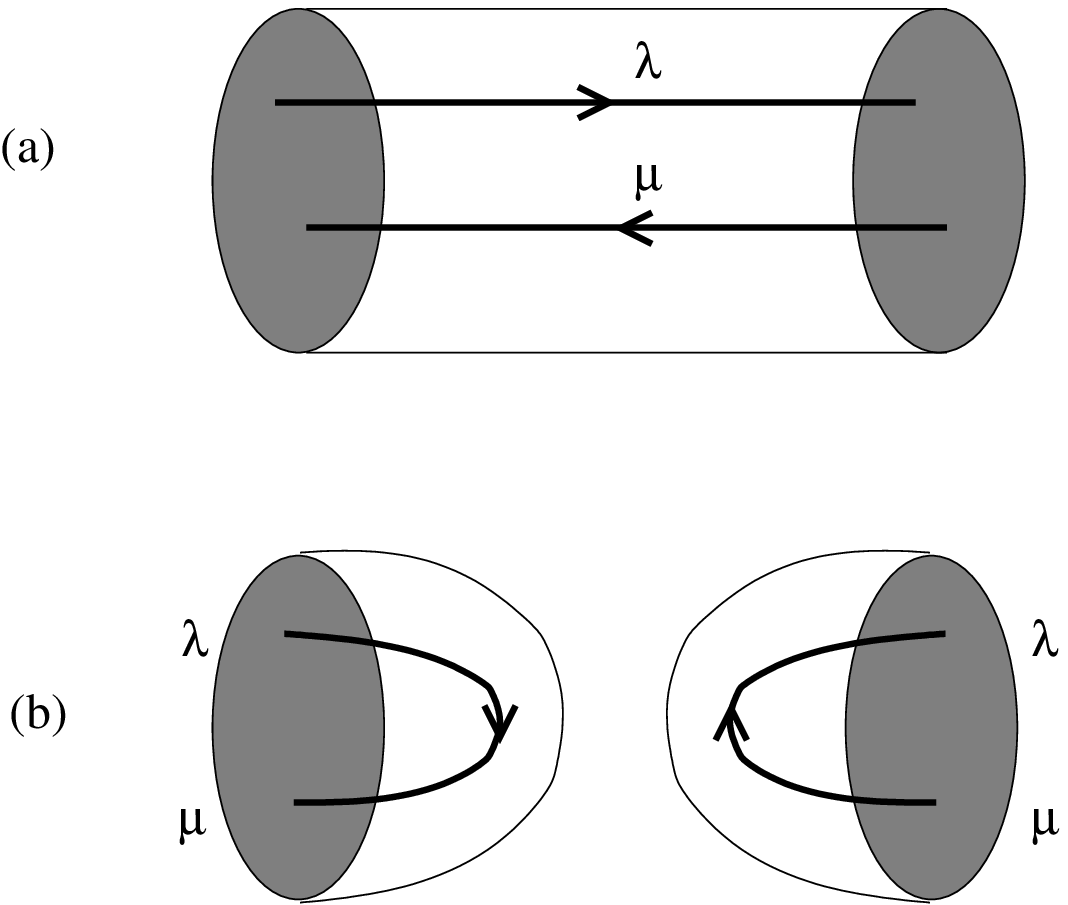}{3}
\par
Since $M$ has two boundary components, the path integral of CS theory
on $M$ yields an element of $\CH_{\Sigma,R_i} \otimes
\CH^*_{\Sigma,R_i}$ where $R_i$ represents the full collection of left
and right moving Wilson line representations.  To go from an element
of this space to a specific CFT correlation function, one must choose
a basis of conformal blocks, or correlators, and expand in this basis.
For example, consider figure~(1) in the case when $\Sigma = S^2$.  In
order for the space of conformal blocks to have nonzero dimension, we
must have $\mu = \lambda^*$ where $\lambda^*$ is the conjugate
representation of $\lambda$. In this case the dimension of
$\CH_{S^2,\{\lambda, \lambda^*\}}$ is one, corresponding to the fact
that there is only one two point function on the sphere, namely
\eqn\CFTcorr{
\langle{\cal O}^\lambda(z_1)\bar{{\cal O}}^{\lambda^*}(\bar{z_2})\rangle = 
\frac{1}{(z_1 - z_2)^{2h_\lambda}},}
where ${\cal O}^\lambda(z)$ is a left moving WZW primary field with conformal
weight $h_\lambda$.  
\par
Now let $\ket{\Psi_M}$ be the CS path integral over the three manifold in
figure (1.a) with $\Sigma=S^2$ and $\mu=\lambda^*$.  Similarly 
let $\ket{\Psi_{S^2}^{\{\lambda,\lambda^*\}}} \otimes 
              \ket{\Psi_{-S^2}^{\{\lambda,\lambda^*\}}}$ be the
path integral in (1.b), and choose $c^2$ times this to be our
standard basis for  
$\CH_{S^2,\{\lambda, \lambda^*\}} \otimes \CH^*_{S^2,\{\lambda, \lambda^*\}}$,
where $c$ is an arbitrary normalization constant.
According to the prescription, we require that $\ket{\Psi_M}$ correspond
to the correlator \CFTcorr, and so its expansion coefficient in the standard basis
should be one:
\eqn\ExpC{
c^2 \,\langle{\Psi_M}  \Big( \ket{\Psi_{S^2}^{\{\lambda,\lambda^*\}}} \otimes 
                      \ket{\Psi_{-S^2}^{\{\lambda,\lambda^*\}}} \Big) = 1.
}
According to the axioms of topological quantum field theory
\refs{\DijkW}, the inner product in \ExpC\ can be computed by gluing
together the two three manifolds in figure (1). After gluing the left (right) side
of figure (1.b) to the right (left) side of (1.a) along the common $S^2$,
the inner product reduces to the path integral of CS theory on a three sphere
with the unknot in the representation $\lambda$. This knot invariant is
given by $S_{0\lambda}$ where $0$ denotes the vacuum representation \refs{\witten}.
Thus we obtain the correct normalization $ c = \frac{1}{\sqrt{S_{0\lambda}}} $.  Indeed
one can readily check that 
$\frac{1}{\sqrt{S_{0\lambda}}} \ket{\Psi_{S^2}^{\{\lambda,\lambda^*\}}}$ has
unit norm in $\CH_{S^2,\{\lambda, \lambda^*\}}$.  This is the CS analogue
of canonical normalization for the chiral primary field ${\cal O}^\lambda(z)$.

\subsec{CS Orbifolds and Boundary CFT's}

\par 
We have seen that the left and right-movers in the full CFT
on a closed surface $\Sigma$ appear in the CS picture at the two
ends of a thickening, $\Sigma \times I$ of $\Sigma$.  Now given that
open strings, or boundary CFT's can be obtained via worldsheet orbifolds
\refs{\horava}, it is natural to try to seek the three dimensional origin
of boundary CFT's in orbifolds of the CS picture.  This was carried
out in detail in \refs{\petrorb}, and here we summarize the results.

\par
Suppose we wish to calculate the bulk one-point function on a Riemann
surface $\Sigma$ with boundary.  Let $\bar\Sigma$ be the oriented
double of $\Sigma$.  By definition $\bar\Sigma$ is a closed, oriented
Riemann surface with an orientation reversing involution $\sigma$ such
that $\Sigma = \bar\Sigma / \sigma$.  The fixed point locus of
$\sigma$ projects to the boundary of $\Sigma$.  For example, when
$\Sigma$ is the disk $D_2$, $\bar\Sigma$ is the sphere $S^2$, and
$\sigma$ is a reflection across a great circle of $S^2$.  Now the key
idea is to consider the thickening $\bar\Sigma \times
I$, and to extend the action of $\sigma$ to a $\ZZ_2$ involution
$\bar\sigma$ on $\bar\Sigma\times I$ which acts as $(z,t) \mapsto
(\sigma(z),1-t)$.  CS theory on the three dimensional orbifold
$M_\Sigma \equiv (\bar\Sigma \times I) / \bar\sigma$ will correspond
to boundary CFT. 
\insfig{2}{(a) $\bar \Sigma \times I$ with $\bar\sigma$ invariant
Wilson line insertions.  The fixed point locus of $\bar\sigma$ carries
an additional Wilson line wrapping the usual insertions. (b) The projection
$M_\Sigma.$ The circular Wilson line in (a) projects to the singular locus
or boundary of $\Sigma$ on the right.}{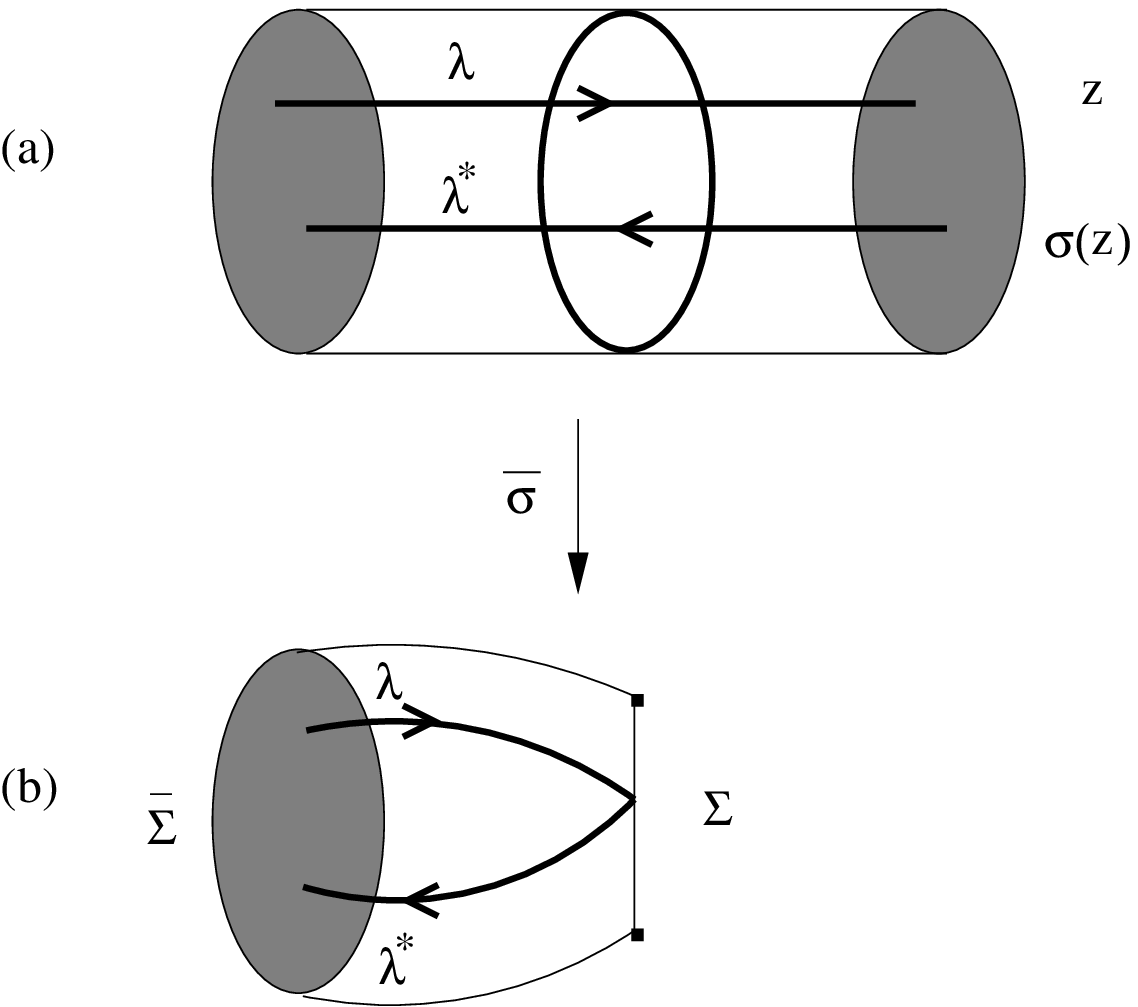}{3}

\par
Now consider the insertion of a left-moving bulk primary field
in the representation $\lambda$ at point $z$ on $\Sigma$.  This point has
two pre-images on $\bar\Sigma$:  $z$ and $\sigma(z)$.  Following the
logic in the previous subsection, we wish to insert Wilson lines
on $\bar\Sigma\times I$ corresponding to the primary field insertion on $\Sigma$,
but now we must respect the $\bar\sigma$ involution.  Thus in addition
to inserting a Wilson line in the representation $\lambda$ oriented from
left to right along $z \times I$, we must also insert another with
the representation $\lambda^*$ oriented from right to left along $\sigma(z) \times I$
as in figure (2.a).  Alternatively one can work directly on the singular
orbifold projection $M_\Sigma$ as in figure (2.b).
\par
Thus we see that the CS orbifold couples the left and right-moving
degrees of freedom together in a manner similar to the way in which
the boundary in CFT reflects the two into each other.  However when
doing CS theory on orbifolds, special attention must be given to the
fixed-point locus of $\bar\sigma$ in $\bar\Sigma \times I$ which gives
rise to singularities in the orbifold $M_\Sigma$.  It was shown in
\refs{\petrorb} that in order to define the path integral, extra data,
namely a choice of holonomies of the CS gauge field around the
singular locus, needs to be given.  Once this choice is made, the
singular locus, which now serves merely as an effective source of
gauge field curvature, can be traded for a link of Wilson lines mimicking this
source.  Thus in figure (2.a), an extra set of Wilson line(s) must
circle the two required by the bulk operator insertion.  The
representation(s) carried by these Wilson lines encode the boundary
condition chosen in the CFT.
\par
For the case when $\Sigma = D_2$, one can visualize $M_\Sigma$ as 
a solid ball with boundary $S^2$ obtained from a continuous deformation of figure 
(2.b) \refs{\FFFSA, \FFFSB}.  One simply contracts all intervals $z \times I$
to a single point, for those $z \in \bar\Sigma$ lying on the fixed point locus of 
$\sigma$.  The manifolds $\Sigma$, $\bar\Sigma$, and $M_\Sigma$
are shown in figure (3), when $\Sigma = D_2$.  More generally it can be shown
that $M_\Sigma$ obeys two important properties: its boundary $\del M_\Sigma$
is $\bar\Sigma$, and
for every pair of images $z$ and $\sigma(z)$ in $\bar\Sigma$, 
there exists a connecting path through the bulk $M_\Sigma$ such that
no two connecting paths for different $z$'s intersect.  These two
properties are apparent in figure (3).  In general, bulk primary field 
insertions on $\Sigma$ in the boundary CFT correspond to Wilson line
insertions along the associated connecting path in $M_\Sigma$.
\insfig{3}{(a) A disk $D_2$ with Cardy boundary condition $\alpha$ and
bulk primary field insertion in the representation $\lambda$. (b) Its oriented
double is $S^2$. The insertion point has two preimages, and the great
circle projects to the boundary.  (c) $M_{D_2}$ is the solid ball, with 
horizontal paths connecting image pairs on the boundary.  Both bulk and
boundary in CFT become linked Wilson lines in three dimensions.}{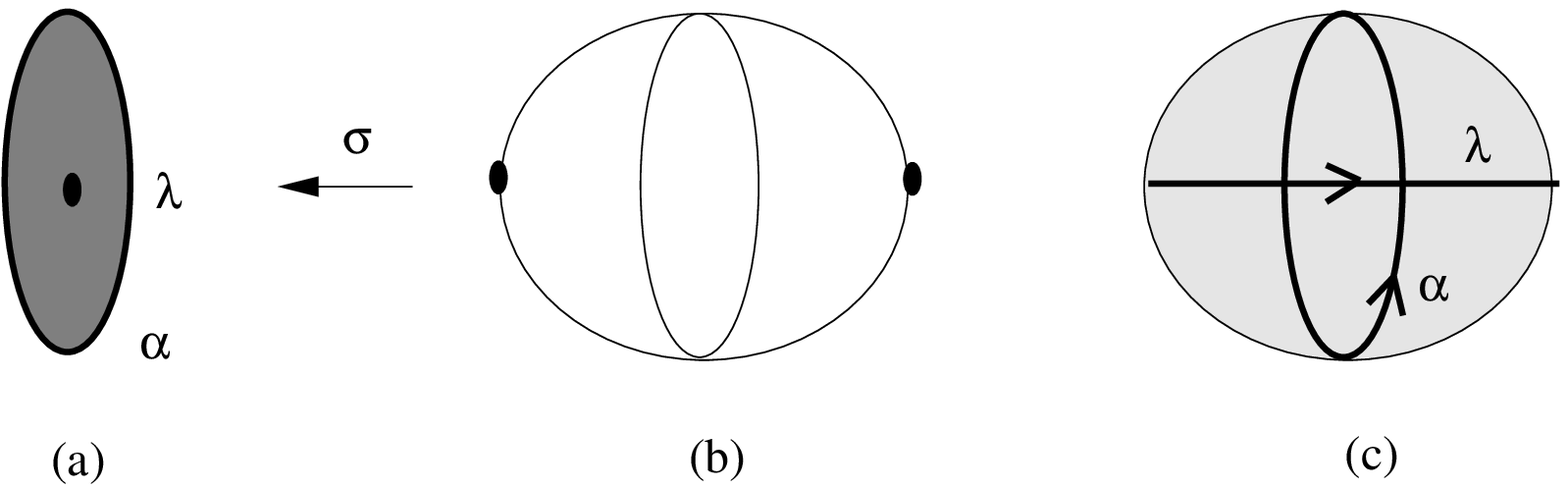}{1.5}

\par
With the above preliminaries, we are now ready to compute the bulk
one-point function on the disk.  It is simply given by the path
integral of CS theory on the solid ball in figure (3.c), and is
therefore an element $\ket{\Psi}$ of the one dimensional vector space
$\CH_{S^2,\{\lambda, \lambda^*\}}$.  We must specify the boundary
condition by specifying the representation of the Wilson line circling
the bulk Wilson line insertion in figure (3.c).  For WZW models, one
can show using the boundary state formalism \refs{\Cardy}, that
symmetry preserving boundary conditions (Cardy states) are in one to
one correspondence with the integrable representations of $\hat{\cal G}$
at level $k$, or equivalently with the set of allowed Wilson line
representations in CS theory.  Thus we simply choose the Wilson line
to be in a representation $\alpha$ corresponding to the boundary condition
we are interested in.
\par
\insfig{4}{There are two ways to get the link in $S^3$ (b). One can either
glue together two solid balls along their boundary $S^2$ as in (a) or
glue together two solid tori via the modular diffeomorphism $S$ as in (c).
 }{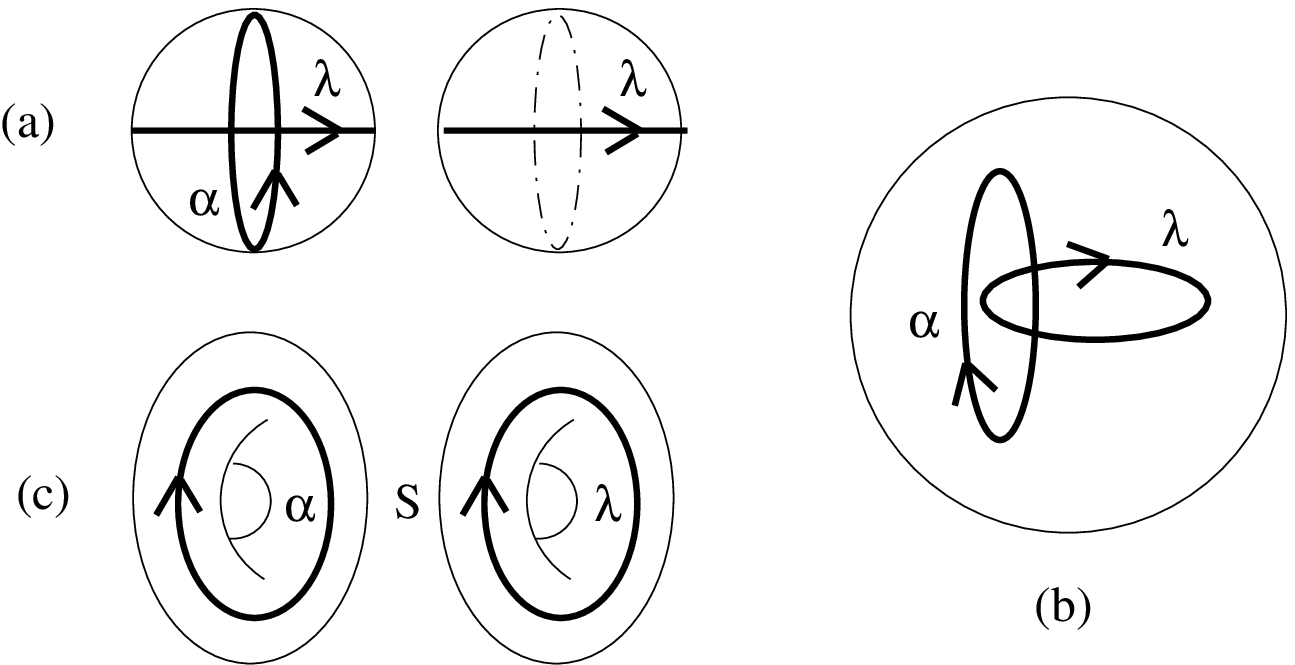}{2}
To compute the actual one-point function, we calculate the inner
product of $\ket\Psi$ with the normalized basis element
$\frac{1}{\sqrt{S_{0\lambda}}}
\ket{\Psi_{S^2}^{\{\lambda,\lambda^*\}}}$.  This involves gluing the
two spheres in figure (4.a) together to obtain a link of Wilson lines
in $S^3$ as shown in (4.b).  The inner product is then given by the
link invariant $S_{\alpha\lambda}$ \refs{\witten}.  One can understand
the appearance of $S_{\alpha\lambda}$ by considering an alternative
surgery to obtain the link invariant.  As shown in figure (4.c) one
can obtain an $S^3$ with a link inside by gluing together two tori
with Wilson lines inside via the gluing map S in \eeS.  As explained
in section 2.1, the path intergrals on the tori in (4.c) yield the
vectors $\ket{\Psi_{T^2}^\lambda}$ and $\ket{\Psi_{T^2}^\alpha}$ in the
space $\CH_{T^2}$.  The result of gluing them together via the map $S$
then computes the matrix element of $S$ in this basis, which is just
the link invariant $S_{\alpha\lambda}$.  Finally, we learn that the
properly normalized one point function $\langle{\cal O}^\lambda(z,\bar{z})\rangle_\alpha$
 of a primary field in the
$\lambda$ representation, in the presence of a Cardy boundary
condition corresponding to the $\alpha$ representation is
$\frac{S_{\alpha\lambda}}{\sqrt{S_{0\lambda}}}$ times the standard one
point function $\langle{\cal O}^\lambda(z,\bar{z})\rangle_0$, recovering 
the results of the boundary state formalism.
\refs{\Cardy}.

\newsec{Boundary Scattering and Aharanov-Bohm: The Kondo Effect}
\par
The Kondo effect \refs{\affleckA-\affleckB} concerns the interaction
of conduction electrons with magnetic spin impurities in metals.
In the case where there is a single impurity of spin $s$ at the origin,
there is a $\delta$-function interaction with the conduction electrons.
Hence in a partial wave expansion, only the S-wave interacts with
the spin, and the problem is essentially one dimensional.  The Hamiltonian
lives on the half-line $r \geq 0$.  In general there may be $k$ flavors
of electrons interacting with the spin $s$ impurity.  A renormalization
group analysis of the interaction Hamiltonian \refs{\affleckA-\affleckB} 
shows that when $2s \leq k$ there exists a nontrivial interacting boundary
CFT that represents the infra-red fixed point of the system.  The physics
of the impurity-electron interactions renormalizes to a universality class
of boundary conditions on the electron density determined by the spin $s$.
\par
In modern language, this fixed point is given by $\SU2{k}$ WZW theory.  The
boundary condition is the Cardy state corresponding to the weight $2s$ 
representation of SU(2).  Furthermore, the single particle scattering
amplitude $S^{(1 \to 1)}$ for the conduction electrons can be written in 
terms of one-point functions of the boundary CFT:
\eqn\scattA{
S^{(1 \to 1)} = \frac{\langle{\cal O}^\lambda(z,\bar{z})\rangle_\alpha}
                {\langle{\cal O}^\lambda(z,\bar{z})\rangle_0}.
}
To make contact with the Kondo problem we choose $\lambda = 1$ corresponding
to the weight of the spin-1/2 representation of the conduction electrons,
and $\alpha = 2s$ corresponding to the weight of the spin-s representation 
of the impurity.
\par
Now using the results at the end of section 2.3, we find simply that the
above ratio of one-point functions yields
\eqn\fans{
S^{(1 \to 1)}  = \frac{S_{\alpha\lambda}}
                     {S_{0\lambda}}.
}
Topologically, this is the ratio of the link invariant in figure (4.b)
to the knot invariant of an unkot.  The boundary condition appears in the numerator,
where it essentially manifests itself as a charged particle traversing the knot
in the denominator.  Physically the one particle boundary scattering amplitude when
viewed from the CS perspective probes the Aharanov-Bohm phase picked up by this
charged particle as it traverses the unknot.
\par
This viewpoint highlights the general structure of boundary scattering
amplitudes in CS theory.  The boundary dependence of the scattering of
quanta is isolated in a bulk one-point insertion on a disk as in
figure (3.a).  In CS language the disk puffs up into a solid ball,
while the field insertion elongates into a Wilson line propagating
across the bulk, as in figure (3.c).  The boundary is promoted to a
charged particle linking the Wilson line.  Finally the phase that the
scattered quanta pick up in the physical theory is none other than the
Aharanov-Bohm phase that the boundary charged particle picks up as it
encircles the propagating Wilson line.

\newsec{Application to a $c=1$ Boundary CFT}

To illustrate the generality of the connection between boundary
scattering and the Aharanov-Bohm effect, we now study a $c=1$ CFT
consisting of a free field interacting with a dynamical, sinusoidal
potential which was analyzed in  \refs{\callan}.  A Wick rotation of
this same theory leads to a hyperbolic potential which was used to
model tachyon condensation in string theory \refs{\sen}.  The
Lagrangian is given by
\eqn\clagrangian{ L =
{1\over 8\pi}\int_0^R d\sigma (\partial_\mu X)^2-
{1\over 2}({g} e^{iX(0)/\sqrt 2}+ {\bar g} e^{-iX(0)/\sqrt 2})~. }
At the self-dual radius this is $\SU2{1}$ WZW theory with (chiral)
currents $J^{\pm}(z) = e^{\pm i\sqrt{2} X(z)}$ and $J^3 = i \del
X(z)$.  The zero modes of these currents, $J^{\pm}$ and $J^3$ are
global $SU(2)_L$ rotations, acting on the Hilbert space.  The key
result in \refs{\callan} is that the boundary state $\ket B$
corresponding to the interaction \clagrangian\ is simply a global
$SU(2)$ rotation of the left-movers relative to the right-movers in
the the original Neumann boundary state $\ket N$.  So if we define
\eqn\ug{
U(g) \equiv e^{i\pi\,(gJ^+ + \bar g J^-)}}
then we have 
\eqn\bnrt{\ket B = U(g) \ket N.} 
\par
Furthermore, in the prescription for calculating scattering amplitudes,
right-moving insertions on the upper half plane (conformally equivalent to the disk)
are mapped to $U(g)$ rotated left-moving insertions on the mirror image
of that point, in the lower half-plane \refs{\callan}.  For $g$ real, $U(g) = e^{i\pi g J_1}$, and
its action on the currents is given explicitly by
\eqn\curract{
\left(\matrix{J^{1} \cr J^{2} \cr J^{3}\cr }\right) 
\mapsto \left(\matrix{1 & 0 & 0 \cr 0 & \cos{(2\pi g)} & \sin{(2\pi g)} \cr
  0 & -\sin{(2\pi g)} & \cos{(2\pi g)} \cr }\right) 
\left(\matrix{J^{1} \cr J^{2} \cr J^{3}\cr }\right).
}
In \bnrt\ and \curract\ both the original boundary state $\ket N$ and the bulk
insertion $\del X$ are acted on by the same rotation $U(g)$.  This means that from
the point of view of CS theory in figure (4.b), both Wilson lines
corresponding to the boundary condition and bulk insertion should be
equally rotated before calculating the Aharanov-Bohm phase associated
to the scattering.  Thus in the calculation of the link invariant in
(4.b) via the surgery (4.c), we first rotate the two vectors in
$\CH_{T^2}$ before gluing them together via the modular $S$ matrix.
For $\SU2{1}$, $\CH_{T^2}$ is spanned by $\ket{\Psi_{T^2}^0}$ and
$\ket{\Psi_{T^2}^1}$ where $0$ and $1$ are the two integrable
representations of $\SU2{1}$.  In the free field language, the
representation $1$ corresponds either to the Neumann boundary
condition or the $\del X$ operator insertion, whereas $0$ corresponds
to a trivial boundary condition or the identity operator.
\footnote{$^\dagger$}{The reader may wonder what happened to the
Dirichlet boundary state. It is actually another SU(2) rotation of the
Neumann boundary state \refs{\Gab}.}  Applying the logic in \fans\ , the single particle
scattering amplitude is then given by the ratio
\eqn\rotscat{
S^{(1 \to 1)} =  \frac{
   \bra{\Psi_{T^2}^1} \tilde U(g)^\dagger S\, \tilde U(g) \ket{\Psi_{T^2}^1}}
  {\bra{\Psi_{T^2}^0} S  \ket{\Psi_{T^2}^1}}.}
The denominator corresponds to turning off the  boundary interaction, but retaining
the bulk insertion.  We have yet to specify the $2 \times 2$ matrix
$\tilde U(g)$ that is the analogue of $U(g)$ in \bnrt\ and \curract.\  A natural
choice is simply to pick the $2$ dimensional representation of the group element
in \ug, namely 
\eqn\ugtwobytwo{
\tilde U(g)= \left(\matrix{\cos{(\pi g)} & i\sin{(\pi g)}\cr 
i\sin{(\pi g)} & \cos{(\pi g)}\cr }\right).}
From equation \eemodS, the explicit form of the modular matrix $S$ at level 
$k=1$ is given by 
\eqn\modsone{
S=\frac{1}{\sqrt{2}} 
\left(\matrix{1 & 1 \cr 1 & -1 \cr} \right).
}
Plugging \ugtwobytwo\ and \modsone\ into \rotscat\ we find the scattering
amplitude
\eqn\tada{
S^{(1 \to 1)} = \cos{2\pi g}
}
which is indeed the result found in \refs{\callan}.  Thus even in this
case, it is correct to interpret this amplitude as an Aharanov-Bohm effect
in three dimensions.  The rotations acting on the vectors in figure (4.c)
can be reinterpreted as a linear combination of Wilson lines running around
the tori via fusion.  The combinations of Wilson lines in the two tori
link each other in $S^3$ after gluing via the modular $S$ matrix and the
combination of phases contributes to the scattering amplitude in \tada.

\par
One may feel slightly uneasy about the ``natural'' choice made in
\ugtwobytwo.  However, it has been shown recently \refs{\Gab}
that all real conformal boundary conditions in $\SU2{1}$ are in
one to one correspondence with group elements $U \in SU(2)$.  The
associated boundary state is simply $U \ket N$, the $U$ rotation of the
Neumann state.  Therefore the space of boundary conditions has a group
structure, and if the CS formula \rotscat\ is to be consistent for this more generic
case, the map from the abstract group element $U$ to a $2\times2$ matrix
$\tilde U$ must be a group homomorphism, or a representation of $SU(2)$. 
Then the identification of $e^{i\pi g J^1}$ with \ugtwobytwo\ is chosen
to enforce \tada.  One can fix the rest of the map
by considering other boundary conditions.  Then \rotscat\ is immediately
generalized to a universal Aharanov-Bohm phase for boundary scattering off
any conformal boundary condition in $\SU2{1}$.

\newsec{Discussion}

\par
From the early days of Kaluza-Klein theory, to the present day study
of $M$-theory, experience has taught us that a higher dimensional
viewpoint can unify and shed insight on various disparate lower
dimensional phenomena.  Certainly CS theory has been able to do this
for CFT on closed Riemann surfaces.  When the CFT correlators are
given by CS path integrals, then the factorization and sewing axiom's
of CFT turn out to be simple consequences of knot theory in 3
dimensions.  By exploiting an orbifolding of this picture, we
elucidated the three dimensional origin of boundary scattering.  Bulk
insertions on a surface $\Sigma$ with boundary become Wilson lines
propagating from one side of a connecting three manifold $M_\Sigma$ to
another.  Boundaries lift to chiral edge states moving along the
orbifold singular locus that encircles these particles.  And the
scattering amplitude is the Aharanov-Bohm phase picked up by these chiral edge
states.
\par
It would be interesting to see how general this picture is and search
for other examples of this correspondence.  Furthermore, the
literature has shown that it pays to take the bulk CS physics in the
connecting three manifold seriously.  For example in \refs{\Ferreira}
it was shown how interactions between the propagating Wilson lines and
non-perturbative instanton processes in $U(1)^n$ CS theory yield the
Narain lattice of toroidial compactification in the CFT.  Thus
non-local effects in the bulk of $M_\Sigma$, such as chiral edge states, and
instanton gases, can affect physics at the boundaries which carry the left
and right-movers of the CFT.   Also there have been long-standing
puzzles about unitarity violations of the boundary scattering amplitude in the Kondo
effect, and the existence of solitonic sectors which restore unitarity.
Perhaps the answers to these 1+1 dimensional puzzles will be found in
the third dimension.

\bigskip
\noindent{\bf Acknowledgement}
\medskip
This work has been supported in part by NSF grant PHY-0244900 and by the 
Berkeley Center for Theoretical Physics. 
\listrefs
\end